\documentclass[9pt,twocolumn,twoside]{osajnl}
\usepackage[ justification=justified]{caption} 
\usepackage{multirow}
\usepackage{titlesec}
\graphicspath{{images/}{../images/}}

\journal{optica} 
\setboolean{shortarticle}{false}
\usepackage{lineno}

\usepackage{subcaption}
\makeatletter
\renewcommand*\subcaption@label{\caption@withoptargs\subcaption@@label}
\makeatother
\usepackage{stfloats}
\usepackage{multirow}
\usepackage{siunitx}
\usepackage{physics}
\DeclareSIUnit\baud{Baud}

\title{Continuous-Variable Quantum Key Distribution at 10~GBaud using an Integrated Photonic-Electronic Receiver}

\author[1,$\dagger$, *]{Adnan A.E. Hajomer}
\author[2,$\dagger$]{C\'{e}dric Bruynsteen}
\author[1,3]{Ivan Derkach}
\author[1]{Nitin Jain}
\author[2]{Axl Bomhals}
\author[2]{Sarah Bastiaens}
\author[1]{Ulrik L. Andersen}
\author[2]{Xin Yin}
\author[1,*]{Tobias Gehring}

\affil[1]{Center for Macroscopic Quantum States (bigQ), Department of Physics, Technical University of Denmark, 2800 Kongens Lyngby, Denmark}
\affil[2]{Ghent University-imec, IDLab, Dep. INTEC, 9052 Ghent, Belgium}
\affil[3]{Department of Optics, Faculty of Science, Palacky University, 17. listopadu 12, 771 46 Olomouc, Czech Republic}
\affil[$\dagger$]{These authors contributed equally}
\affil[*]{Corresponding authors: aaeha@dtu.dk, tobias.gehring@fysik.dtu.dk}

\begin{abstract}
Quantum key distribution (QKD) is a well-known application of quantum information theory that guarantees information-theoretically secure key exchange. As QKD becomes more and more commercially viable, challenges such as scalability, network integration, and high production costs need to be addressed. Photonic and electronic integrated circuits that can be produced in large volumes at low cost hold the key to large-scale deployment of next-generation QKD systems.
Here, we present a continuous-variable (CV) QKD system using an integrated photonic-electronic receiver that combines a silicon photonic integrated circuit implementing a phase-diverse receiver with custom-designed GaAs pHEMT transimpedance amplifiers. The QKD system operates at a classical telecom symbol rate of \SI{10}{\giga\baud}, generating high secret key rates exceeding {\SI{0.7}{Gb\per\second}} over a distance of \SI{5}{\km} and {\SI{0.3}{Gb\per\second}} over a distance of \SI{10}{\km}. The secret keys are secure against collective attacks with finite-size effects taken into account. Well-designed digital signal processing enabled the high-speed operation. 
Our experiment sets a new record for secure quantum communication and paves the way for the next generation of CV-QKD systems.
\end{abstract}

\setboolean{displaycopyright}{false}

\begin{document}

\maketitle
\section{Introduction}
Quantum key distribution (QKD) is a method for sharing cryptographic keys between remote users based on the laws of quantum mechanics~\cite{BEN84,ekert1991quantum}. When combined with one-time pad encryption, QKD offers information-theoretic secure data transmission that current or future technologies cannot break. However, this requires a shared secret key that can only be used once, with a length of at least as long as the message~\cite{shannon1949communication}. Therefore, increasing the secret key rate (SKR) of QKD is crucial for enabling secure communication in networks with a large number of users or high data-rate applications, such as distributed storage encryption and high-speed access networks~\cite{diamanti2016practical,sasaki2017quantum}.   

 In the continuous variable (CV) flavour of QKD, quantum information is encoded in the continuous degrees of freedom of quantum systems, such as the amplitude and phase quadrature of the electromagnetic field of light~\cite{grosshans2002continuous}. This protocol has gained significant attention in the scientific community due to its ability to provide high rates. It can approach the ultimate limit of secure communication known as the Pirandola-Laurenza-Ottaviani-Banchi (PLOB) ~\cite{pirandolaFundamentalLimitsRepeaterless2017} bound. In the prepare-and-measure version of the CV-QKD protocol, the sender, Alice, prepares coherent quantum states using a quadrature modulator and sends them to the receiver, Bob. The quantum states propagate through an insecure channel controlled by a potential eavesdropper (Eve). Bob measures the quantum state using coherent detection, e.g., heterodyne or homodyne detection, facilitated by a local oscillator (LO)~\cite{weedbrook2012gaussian,kikuchi2015fundamentals}. These detection techniques have to be quantum-noise-limited across the entire quantum signal bandwidth. 

Numerous high-rate CV-QKD systems have been developed recently, using both Gaussian~\cite{jainPracticalContinuousvariableQuantum2022, sarmientoContinuousvariableQuantumKey2022,ren2021demonstration,wang2018high,huang2015continuous,wang201525,wang2020high, tian2023high} and discrete modulation ~\cite{roumestanExperimentalDemonstrationDiscrete2022, wangSubGbpsKeyRate2022,pan2022experimental,milovanvcev2020spectrally,eriksson2020wavelength} of coherent states. While Gaussian modulation-based protocols have the most advanced security proof~\cite{jainPracticalContinuousvariableQuantum2022}, they also require a high-resolution digital-to-analog converter (DAC) to approximate the analog constellation space adequately. This is problematic for increasing the symbol rate of the QKD system since the signal-to-noise performance of DACs drops off at high sampling rates \cite{adc_survey}. On the other hand, discrete modulation uses a discrete constellation space, making it much more compatible with high-speed wireline components. However, thus far, there are hardly any discrete-modulated (DM) CV-QKD systems operating in the multi-GBaud regime, which is due to the difficulty in ensuring consistent and quantum-noise-limited behavior over such large bandwidths. 

Although CV-QKD is well-suited for photonic integration thanks to its compatibility with standard telecom components \cite{aldamaIntegratedQKDQRNG2022}, recent efforts have focused primarily on integrating discrete variable (DV) QKD \cite{zhengHeterogeneouslyIntegratedSuperconducting2021, beutelDetectorintegratedOnchipQKD2021}. While Zhang \emph{et al}. \cite{zhangIntegratedSiliconPhotonic2019b} have been able to demonstrate the feasibility of using integrated circuits for CV-QKD, their achieved SKR of \SI{0.25}{Mb\per\second} is relatively low, especially compared to bulk fiber-based CV-QKD implementations~\cite{roumestanExperimentalDemonstrationDiscrete2022, wangSubGbpsKeyRate2022,pan2022experimental,milovanvcev2020spectrally,eriksson2020wavelength,tian2023high}. In this work, we show that integrating CV-QKD not only offers the potential for miniaturization and cost-effectiveness but also provides higher key rates thanks to high-bandwidth components and low-noise design techniques. Therefore, integrating CV-QKD has the potential to significantly enhance the cost-effectiveness, performance, and practicality of QKD systems.

In this article, we report our efforts to overcome the bandwidth limitations in CV-QKD. Specifically, we focus on improving the receiver's performance by introducing a co-integrated phase diverse receiver consisting of a silicon photonics optical front-end and custom-integrated transimpedance amplifiers designed in a 100 nm GaAs pHEMT technology. By utilizing high-speed integrated components and implementing a low-noise design, the receiver has access to a shot-noise-limited bandwidth of more than 20 GHz. 

At the transmitter's side, we design a digital signal processing (DSP) pipeline, including a pre-emphasis filter, for quantum state preparation. This enhancement allows our transmitter to operate at a symbol rate of \SI{10}{\giga\baud}. By combining the improvements in the receiver and transmitter, we demonstrate the highest secret key rate (SKR) of any DM coherent state CV-QKD system to date. Specifically, we achieve SKRs of {\SI{0.92}{Gb\per\second}} and {\SI{0.48}{Gb\per\second}} in the asymptotic regime and {\SI{0.737}{Gb\per\second}} and {\SI{0.315}{Gb\per\second}} considering finite-size effects over distances of \SI{5}{\km} and \SI{10}{\km}, respectively. These advances represent a significant step towards realizing practical, high-performance CV-QKD systems.

\section{Integrated receiver}

\begin{figure*}
\begin{minipage}[c]{0.49\linewidth}
    \centering
    \includegraphics[width=\textwidth]{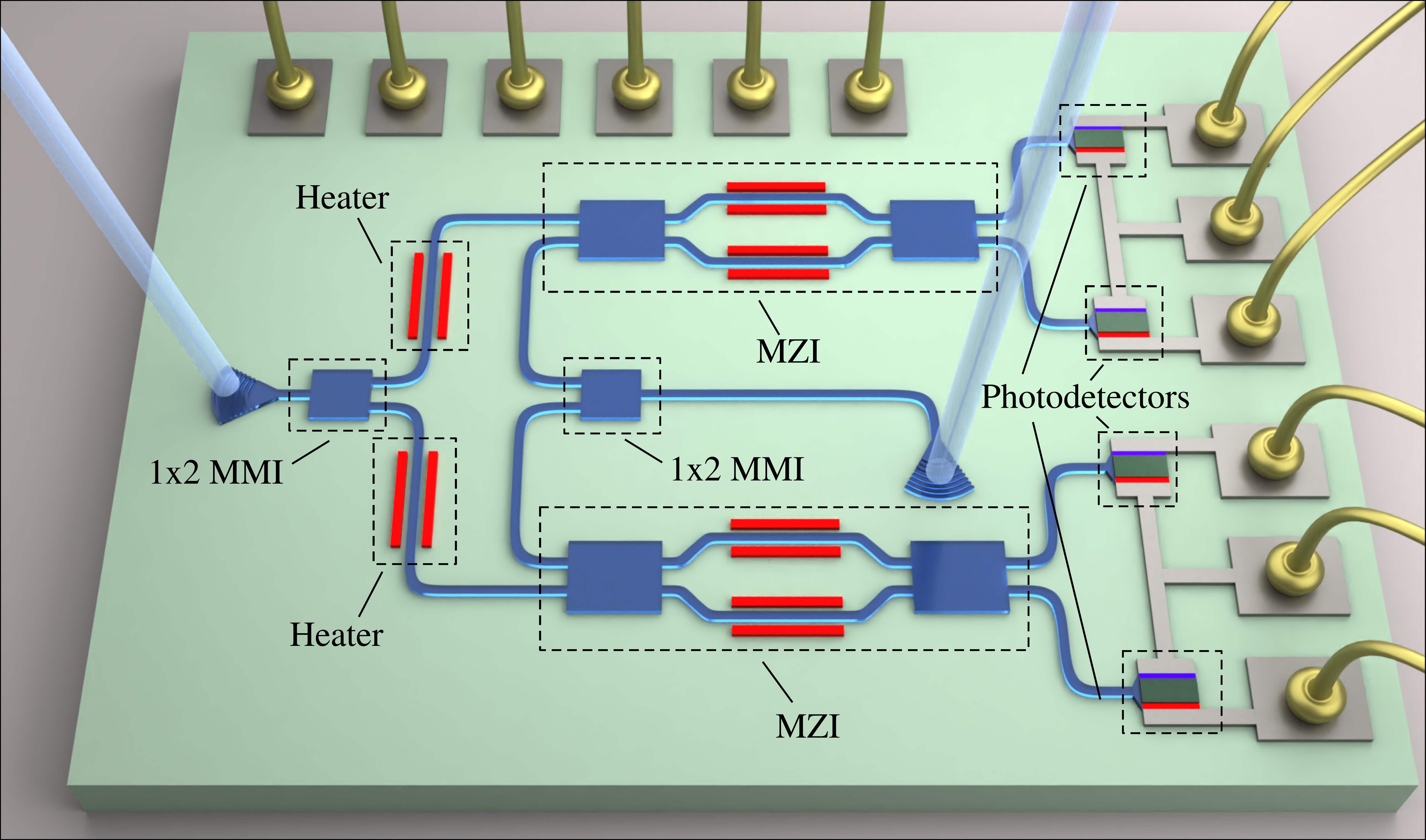}
    \caption{Diagram of the photonic IC. MZI: Mach-Zehnder interferometer; MMI:  multimode interferometer}
    \label{fig:pic_diagram}
\end{minipage}
\hspace{0.01\textwidth}
\begin{minipage}[c]{0.49\linewidth}
    \centering
    \includegraphics[width=0.6\textwidth]{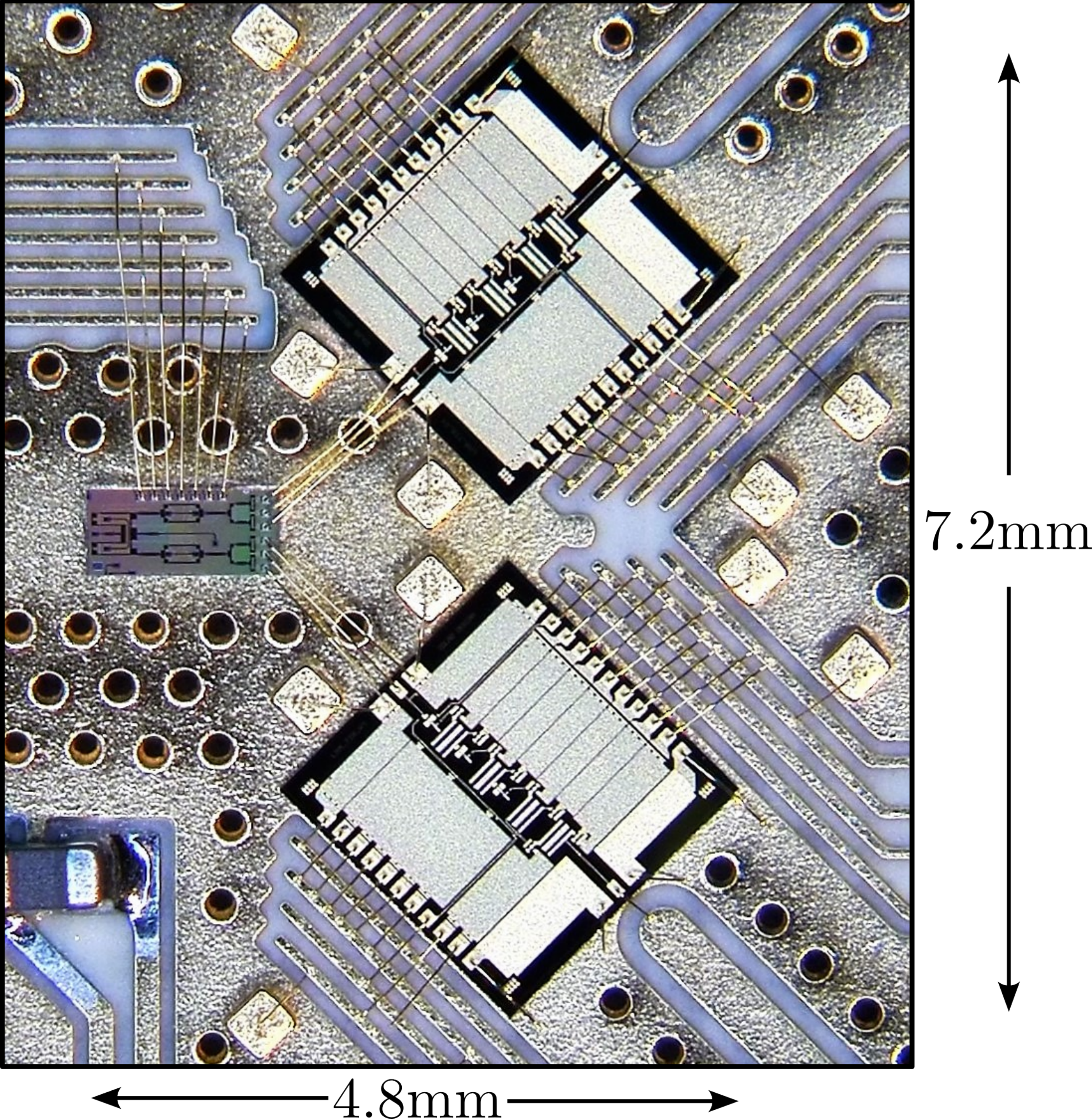}
    \caption{Micrograph of the integrated receiver assembly}
    \label{fig:micrograph}
\end{minipage}
\end{figure*}

Our phase-diverse optical front-end, designed using imec's iSiPP50G silicon photonics platform~\cite{imecabsil}, is shown schematically in Fig.~\ref{fig:pic_diagram}. Light was coupled into the chip using two grating couplers, with one input receiving the LO and the other receiving the incoming quantum signal. The quantum signal and the LO were split into two arms using a 1x2 multimode interferometer (MMI) to measure the conjugate quadratures. Two waveguide heaters were added in the LO path and actively controlled to ensure a $90 ^\circ$ phase shift between the measured quadratures. This was followed by two balanced homodyne detectors, which were composed of a Mach-Zehnder interferometer (MZI) and two photodetectors with a high responsivity of \SI{1.1}{A \per W} \cite{imecabsil}. The MZI was implemented using two 2x2 MMIs and two heaters. The heaters were tuned using feedback from the transimpedance amplifier (TIA) to minimize the amount of direct current (DC) flowing into the TIA which increases the common-mode rejection ratio. The overall measured efficiency of the phase-diverse homodyne receiver was $\eta = 44\%$, with most of the loss attributed to the grating coupler (insertion loss (IL) = 2.5 dB \cite{imecabsil}). However, this may be improved in future designs by employing edge couplers (IL = 1.1 dB \cite{heVGrooveAssistedPassive2019}) to achieve a theoretical efficiency of $\eta \approx 61\%$.

Each pair of balanced photodetectors was connected to a separate TIA, designed in a 100nm GaAs pHEMT process. The TIA implemented a three-stage core amplifier, enabling higher transimpedance values and lowering the electronic noise \cite{bruynsteenIntegratedBalancedHomodyne2021}. A micrograph of the PIC with the two TIA dies is shown in Fig.~\ref{fig:micrograph}. The overall size of the three chips was \SI{4.8}{\mm} x \SI{7.2}{\mm}. The chips were mounted on an interposer printed circuit board (PCB) which was temperature stabilized by a thermo-electric cooler and a surface-mounted thermistor. The output of each TIA was connected to a \SI{50}{\ohm} transmission line, which was terminated in a coaxial connector.

\begin{figure*}[h]
    \centering
    \includegraphics[width = \linewidth]{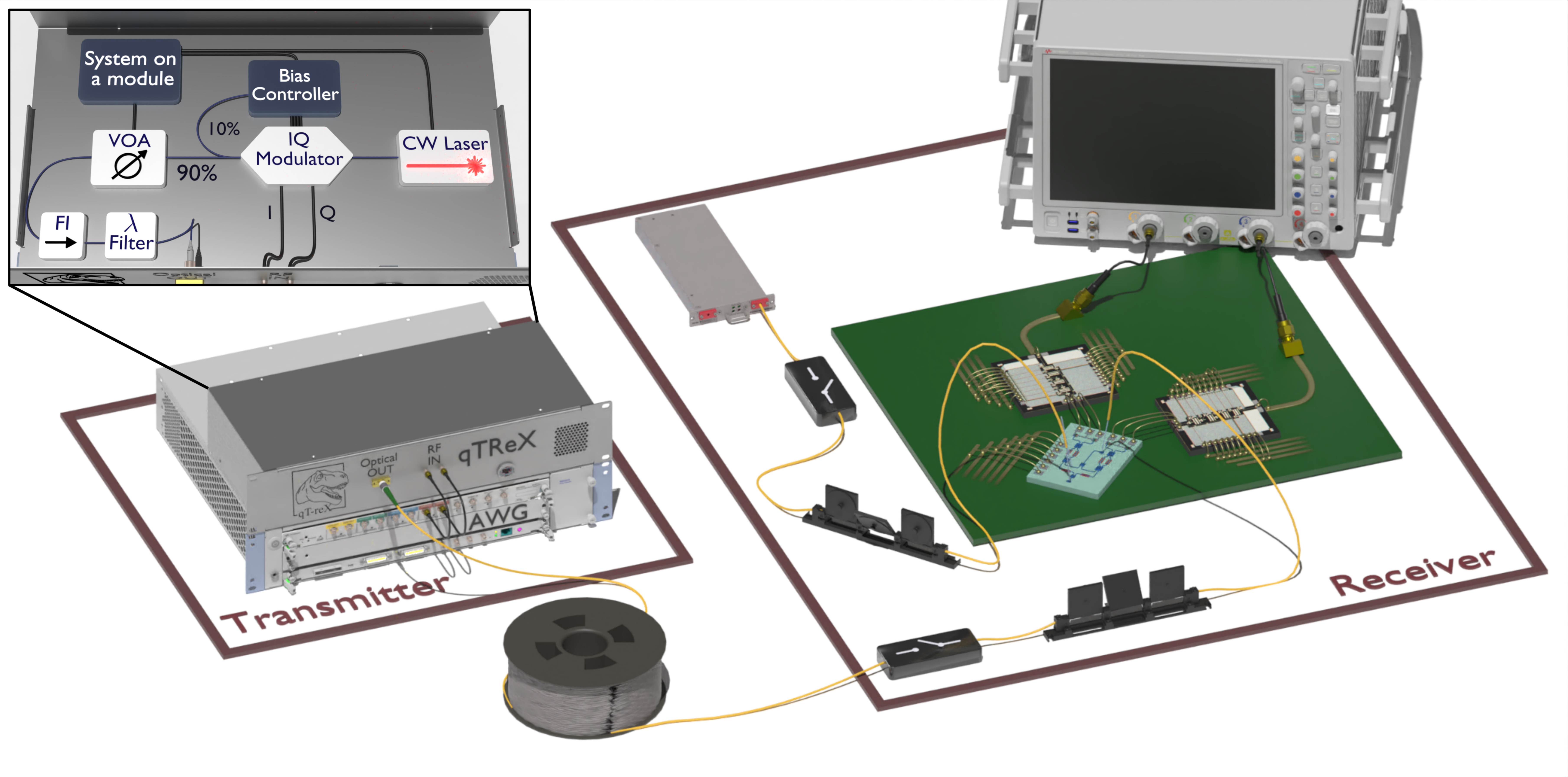}
    \caption{\textbf{High-rate CV-QKD set-up}. Diagram of the QKD system with all key components. At the transmitter side: CW laser (continuous wave laser), IQ modulator (in-phase and quadrature modulator), VOA (variable optical attenuator), FI (Faraday isolator), wavelength filter and AWG (arbitrary waveform generator). At the receiver side: CW laser, two optical switches, two polarization controllers, the photonic and electronic integrated circuits on an interposer PCB and RTO (real-time oscilloscope).}
    \label{fig:setup_diagram}
\end{figure*}

\section{High-rate CV-QKD SYSTEM}
Figure~\ref{fig:setup_diagram} shows the schematic of our high-speed CV-QKD system, which consisted of two stations: the transmitter (Alice) and the receiver (Bob), connected through a quantum channel made of standard single-mode fiber (SSMF). The system was designed to work semi-autonomously, meaning that it was able to perform state preparation and measurement without user intervention. This was achieved by a single system on a module that monitored and controlled the different optical and electronic sub-components. In the following subsection, we will provide a detailed description of each station. 
\subsection{Transmitter (Alice)}
Alice's station comprised a high-speed, 8-bit arbitrary waveform generator (AWG, Keysight M8195A)
with two channels operating at a sampling rate of 32 GSample/s driving our developed transmitter unit, known as qTReX \cite{jainQTReXSemiautonomousContinuousvariable2022}, housing all optical and electronic components. The inset in Fig.~\ref{fig:setup_diagram} shows the internal schematic of qTReX. A 1550 nm continuous wave (CW) laser with a linewidth of 100 Hz was used as an optical source. The coherent states were generated in the sideband frequencies by modulating the CW laser using an in-phase and quadrature (IQ) modulator driven by the AWG. The IQ modulator was biased in its minimum transmission point by controlling the DC voltages using an automatic bias controller~\cite{hajomer2022modulation}. The modulation variance of the generated thermal state was controlled by a variable optical attenuator (VOA). To avoid Trojan-horse attacks enabled by back reflection, a Faraday isolator (FI) and wavelength filter were connected to the output of the VOA.

Producing coherent states at a GBaud symbol rate requires a transmitter with a large bandwidth. In practice, however, any transmitter exhibits a high-frequency roll-off, introducing correlations (intersymbol interference (ISI)) between the transmitted symbols. From a security point of view, this can destroy the independent and identically distributed property of the quantum symbols, which would violate an assumption commonly made in security proofs~\cite{laudenbach2018continuous}. 

To cope with this issue, we carefully designed a digital signal processing (DSP) pipeline, including a pre-emphasis filter ~\cite{rafique2014digital, Proakis2007}, for coherent state preparation, as depicted in Fig.~\ref{fig:DSP}. The complex amplitude of the coherent state $\ket{\alpha}=\ket{x+jp}$ was drawn from a probabilistically shaped discrete constellation with a defined modulation order (M) and a Gaussian-like probability distribution. This is also known as probabilistic constellation shaping (PCS) in classical telecommunication~\cite{Proakis2007}. These coherent states were drawn at symbol rates of \SI{8}{\giga \baud} or \SI{10}{\giga \baud} and up-sampled to 32 GSample/s, after which they were pulse-shaped using a root-raised cosine filter with a roll-off of 0.2. To compensate for the high-frequency slope of the transmitter, the baseband signal was digitally pre-emphasized using the inverse frequency response of the transmitter. The inset of Fig.~\ref{fig:DSP} shows the frequency response of the transmitter, including the IQ modulator, AWG and the RF cables, as well as the spectrum of the quantum signal before and after the pre-emphasis filter. To share a reference phase between the transmitter and the receiver, pilot tones at 8 GHz or 7 GHz were frequency multiplexed with the 10 GBaud or 8 GBaud pre-emphasized quantum signals, respectively.

\subsection{Receiver (Bob)}
To measure the coherent quantum states, intradyne detection~\cite{kikuchi2015fundamentals} was performed using our integrated phase-diverse receiver and a free-running CW laser with respect to the transmitter's laser. For intradyne detection, the frequency difference between these lasers was set to be less than half of the bandwidth of the quantum signal. To maximize the coupling into the chip, two polarization controllers (PCs) were used to adjust the polarization of the LO and the quantum signal. For autonomous system calibration, i.e., for vacuum noise measurements (Signal off and LO on) and electronic noise measurements (Signal off and LO off), electronically-controlled switches were added in the LO and signal path. After the phase-diverse receiver, the signal was digitized using an 8-bit real-time oscilloscope (RTO, Keysight DSA Z634A) with a sampling rate of 80 GSamples/s, which was clock synchronized with the AWG using a 100 MHz reference clock. 

\begin{figure*}
\centering
\includegraphics[width=\linewidth]{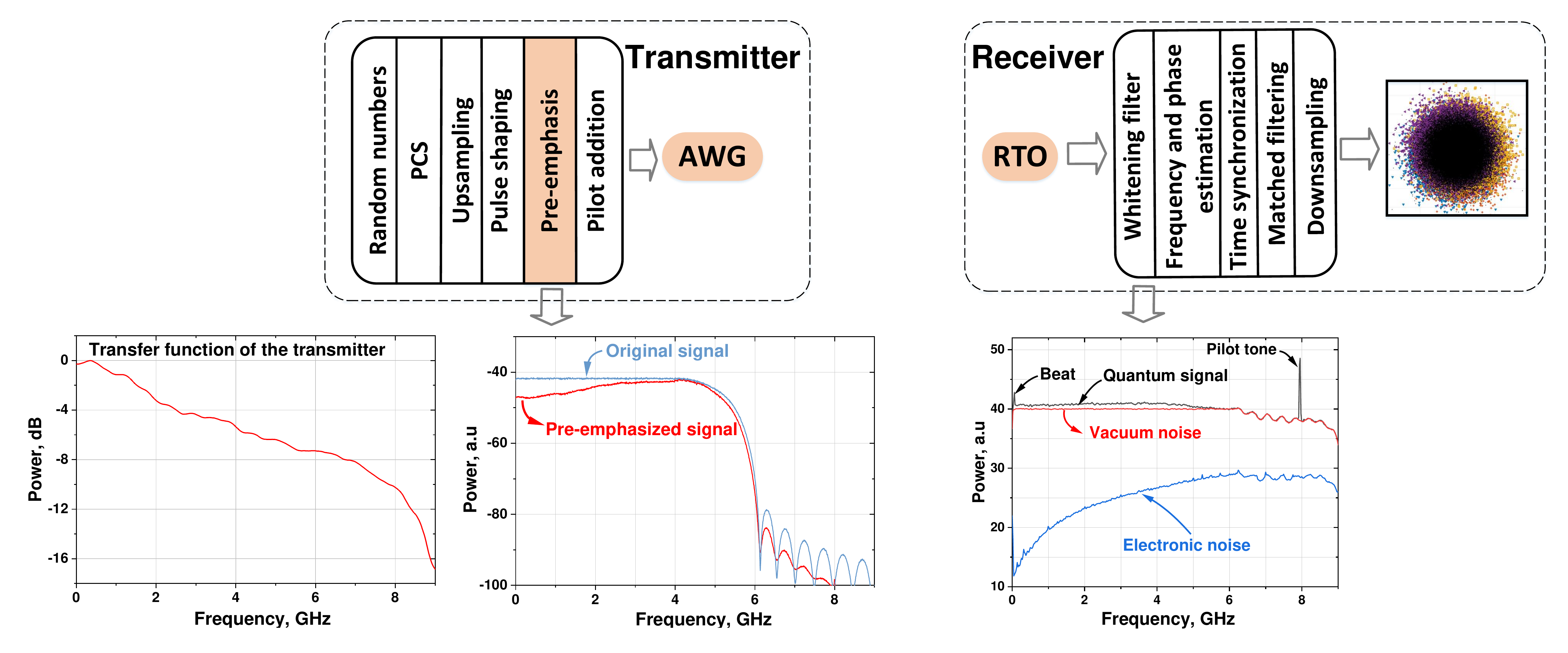}
\caption{\textbf{The DSP routine of GBaud CV-QKD system}. AWG: arbitrary waveform generator; PCS: probabilistic constellation shaping; RTO: real-time oscilloscope. See the main text for the details.} 
\label{fig:DSP}
\end{figure*}

Figure~\ref{fig:DSP} shows the DSP routine used for quantum symbols recovery. First, a digital post-equalizer (whitening filter) was applied to the quantum signal, the vacuum noise and the electronic noise traces to compensate for the high-frequency roll-off of the receiver. The filter coefficients were obtained by fitting the inverse frequency response of the receiver, computed from the vacuum noise measurement. The equalized spectrum is depicted in the left inset of Fig.~\ref{fig:DSP}. Next, the pilot tone frequency was estimated using a Hilbert transform on the filtered pilot tone and the linear fit of its phase profile. To extract the relative phase between the transmitter and the receiver, the pilot tone was baseband transformed using the pilot frequency estimate. After that, the quantum signal was baseband transformed by the frequency difference between the known pilot frequency at the transmitter and the pilot frequency estimate at the receiver. The phase of the quantum signal was then corrected using the extracted phase of the pilot tone. Afterwards, the temporal shift due to the propagation delay of the quantum channel and different electronic components was obtained by calculating the cross-correlation between the reference transmitted samples and the received samples. The corresponding quantum symbols were obtained after root-raised-cosine matched filtering and downsampling. Finally, the quantum symbols were rotated to compensate for a residual phase shift due to the frequency difference between the quantum signal and the pilot tone.


\section{Security analysis} 

Contrary to Gaussian CV-QKD protocols where coherent states follow a two-dimensional zero-centered continuous Gaussian distribution $\mathcal{N}(0,\Sigma)$, in discrete-modulated (DM) CV-QKD protocols, coherent states are drawn from a discretized set $\{\alpha_k\}_{k=1,\dotsc M}$, with respective probabilities $\mathcal{P}(\alpha_k)$, where $M$ is the cardinality of the discrete constellation. On the receiver side the measurement of incoming quantum states yields a complex number $\zeta_k$. After transmitting and receiving a block of $N$ symbols, the trusted parties hold two correlated strings (one for each quadrature) of equal length, which they then correct for errors and use to distill secret keys, i.e., identical random sequences that are completely uncorrelated with any unauthorized party.

The security analysis of QKD protocols is based on the equivalent entanglement-based representation of state preparation. In this representation, Alice generates an entangled state $|\Psi\rangle_{AB}$ and measures one of the modes to conditionally prepare the other. Attribution of the channel to Eve, without loss of generality, implies that after transmission Eve uses her share of the joint pure state $|\Psi\rangle_{ABE}$ to infer Bob's measurement outcomes.

The security against collective attacks is defined as the positivity of accessible information difference \cite{devetak2005distillation}:
\begin{equation}
    R_\infty=\beta I_{AB}-\chi_E,
    \label{eq:key}
\end{equation}
where $\beta$ is the reconciliation efficiency that indicates the inability to recover full mutual information $I_{AB}$ between Alice's and Bob's data sets, while $\chi_E$ is the Holevo bound that bounds the information an adversary may hold on the generated data sequence at the reference side (Bob). Both $I_{AB}$ and $\chi_E$ can be evaluated based on the covariance matrix $\Gamma_{AB}$ of the state shared after channel $\rho_{AB}$. While Alice and Bob can directly obtain variances of local quadrature operators, the term corresponding to the correlations between modes $A$ and $B$ is not trivially assessed when a general non-Gaussian measurement of the former prepares the state in the latter. We employ a theory developed by A.
Denys, P. Brown, and A. Leverrier \cite{denys2021explicit} that allows to analytically bound the correlation term in $\Gamma_{AB}$ and consequently the secure key rate (SKR) in Eq. (\ref{eq:key}) in the asymptotic regime of infinitely many exchanged quantum states.

\begin{figure}[!t]
    \centering
    \includegraphics[width = .99
\linewidth]{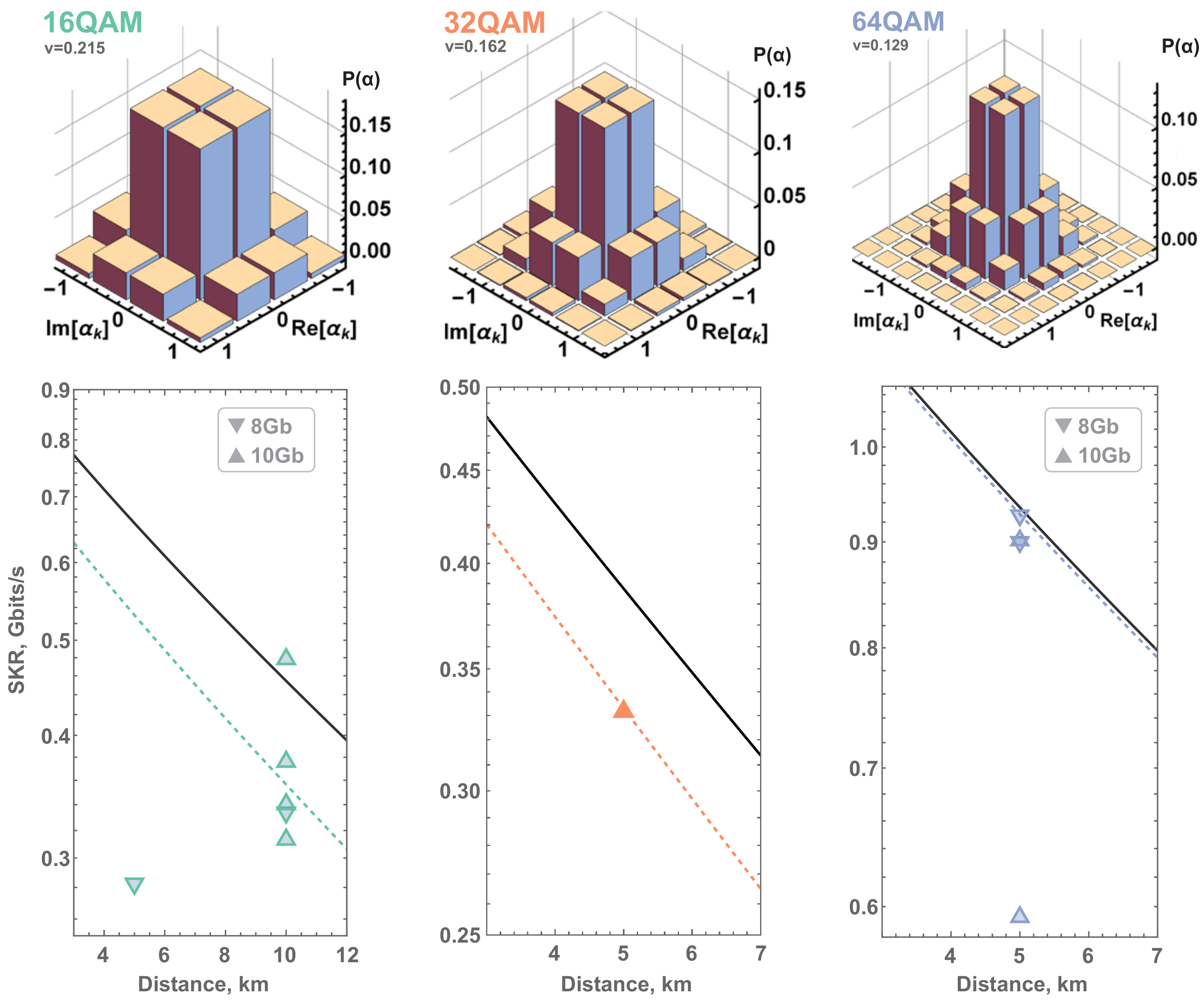}
    \caption{(\textbf{Top}) The constellation probability distributions used in the experiment: 16, 32 (four corner points have a probability of 0), and 64 QAM. $\alpha$'s are in $\sqrt{\text{SNU}}$. (\textbf{Bottom}) Asymptotic secure key rate against fiber length (\SI[per-mode=symbol]{0.2}{\dB\per\km}). Lines show the theoretical curves for GG02 and respective constellation with a repetition rate of \SI{10}{\giga\baud}, $\beta=95$\%, modulation variance $V_M$ excess noise $\epsilon$ and coupling efficiency are mean of observed experimental values for respective constellation. Points show the experimental results for each constellation and distance used, with repetition rate of \SI{8}{\giga\baud} (down triangle) and \SI{10}{\giga\baud} (up triangle). }
    \label{fig:fig5}
\end{figure}

We use three constellations for quadrature amplitude modulation (QAM) given by discretized Gaussian distributions: $M=16,\,32,\,64$ (see top Fig. \ref{fig:fig5}). Each constellation is characterized by the variance of the distribution $V_M$ and assigned probabilities within the constellation $\mathcal{P}(\alpha_k)$, determined by the cardinality $M$ and the parameter $\nu$ related to the variability or statistical dispersion of the probability distribution. The optimal value of modulation variance $V_M=2\alpha^2$ that maximizes the SKR strongly depends on the post-processing efficiency $\beta$. We choose the variance $V_M\approx 1$ as it allows us to operate at low excess noise and it ensures that the mutual information $I_{AB}$ of the continuous-Gaussian protocol is a good approximation for the DM protocol~\cite{wu2010impact}.
The choice of parameter $\nu$ is crucial for the protocol's performance and thus was optimized based on the cardinality $M$ and modulation variance $V_M$. \par
In the analysis, we presume that the receiver station is fully trusted. Hence, the heterodyne detector efficiency $\eta=44\%$ and electronic noise with variance $V_{el}$ were modeled as linear coupling of the signal to a thermal noise source purified by trusted parties. This partially decouples Eve from Bob's accumulated key and decreases the Holevo bound $\chi_E$ \cite{usenko2016trusted}. \par
\begin{table*}[t]
\caption{Summary of experimental parameters and secure key rates for best results at each distance. Reconciliation efficiency $\beta=0.95$, detector efficiency $\eta=0.44$, block size $N=1.6\times10^7$.} 
    \centering
    \begin{tabular}{|l|l|l|l|l|l|l|l|l|l|l|}
\hline
M  & $\nu$, a.u. & \begin{tabular}[c]{@{}l@{}}Rep. rate\\ s, GBaud\end{tabular} & \begin{tabular}[c]{@{}l@{}}Fiber\\ d, km\end{tabular} & $V_{M}$, SNU & $T$, a.u. & $V_{el}$, \% SNU & $\epsilon$,  \% SNU & \begin{tabular}[c]{@{}l@{}}$\text{R}_\infty$, \\ bits/symbol\end{tabular} & \begin{tabular}[c]{@{}l@{}}$\text{R}_\text{finite}$, \\ bits/symbol\end{tabular} & \begin{tabular}[c]{@{}l@{}}$\text{SKR}_\text{finite}$, \\ Gbits/sec\end{tabular} \\ \hline
16 & 0.215 & 10                                                            & 10                                                    & 0.87         & 0.569        & 6.50             & 2.622                & 0.048                                                                      & 0.035                                                                             & 0.351                                                                            \\ \hline
16 & 0.215 & 8                                                            & 5                                                     & 1.01         & 0.618        & 4.95             & 5.187               & 0.035                                                                      & 0.021                                                                             & 0.171                                                                            \\ \hline 

32 & 0.162 & 10                                                           & 5                                                     & 0.93         & 0.702        & 6.76             & 7.183               & 0.033                                                                      & 0.019                                                                             & 0.194                                                                            \\ \hline
64 & 0.129 & 8                                                            & 5                                                     & 1.03         & 0.733        & 5.03             & 1.590               & 0.115                                                                      & 0.093                                                                             & 0.746                                                                            \\ \hline
    \end{tabular}

\label{tab:results}
\end{table*}
\section{Results}
The results of the security analysis in the asymptotic regime are shown in Fig.~\ref{fig:fig5} (bottom).  The triangles show experimental results for several measurements taken over distances of \SI{5}{\km} and \SI{10}{\km} and symbol rates of \SI{8}{\giga\baud} and \SI{10}{\giga\baud} for the three modulation formats. Assuming an information reconciliation efficiency of $95\%$, with 64~QAM the highest secret key rate of 0.92 Gb/s was achieved over the \SI{5}{\km} fiber channel due to the high cardinality and the low excess noise compared to 16~QAM and 32~QAM. For the measurement set with $M=32$, a higher measured excess noise led to a smaller SKR of {{\SI{0.33}{Gb\per\second}}}, which is comparable to results for lower cardinality $M=16$ set (in the same fiber) with lower excess noise. Table ~\ref{tab:results} shows a summary of the best experimental results for each constellation and distance.

Furthermore, the lines of  Fig.~\ref{fig:fig5} (bottom) show the theoretical predictions for protocols at $V_M=1$ SNU with $M=16,\,32,$ and $64$ (with optimized $\nu$), and GG02 protocol with Gaussian modulation~\cite{grosshans2002continuous}. Theoretical predictions assume (from left to right) mean modulation variance $V_M=0.87,\,0.93,\,1.02$ SNU, fiber loss of $0.2$ dB/km and mean coupling efficiency of $\eta_D= 84.5,\,88.4,\,92.3 \%$, the mean excess noise observed in our experiment $\epsilon=0.035,\,0.071,\,0.032$ SNU (at channel input) and the mean electronic noise $V_{el}=0.061,\,0.067,\,0.054$ SNU and highest repetition rate $s$ of \SI{10}{\giga\baud}, with the secure key given as $s\times R_\infty$. Using 64 constellation points allows to essentially recover the performance of the protocol with ``continuous'' Gaussian modulation.

\begin{figure}
    \centering
    \includegraphics[width = .8
\linewidth]{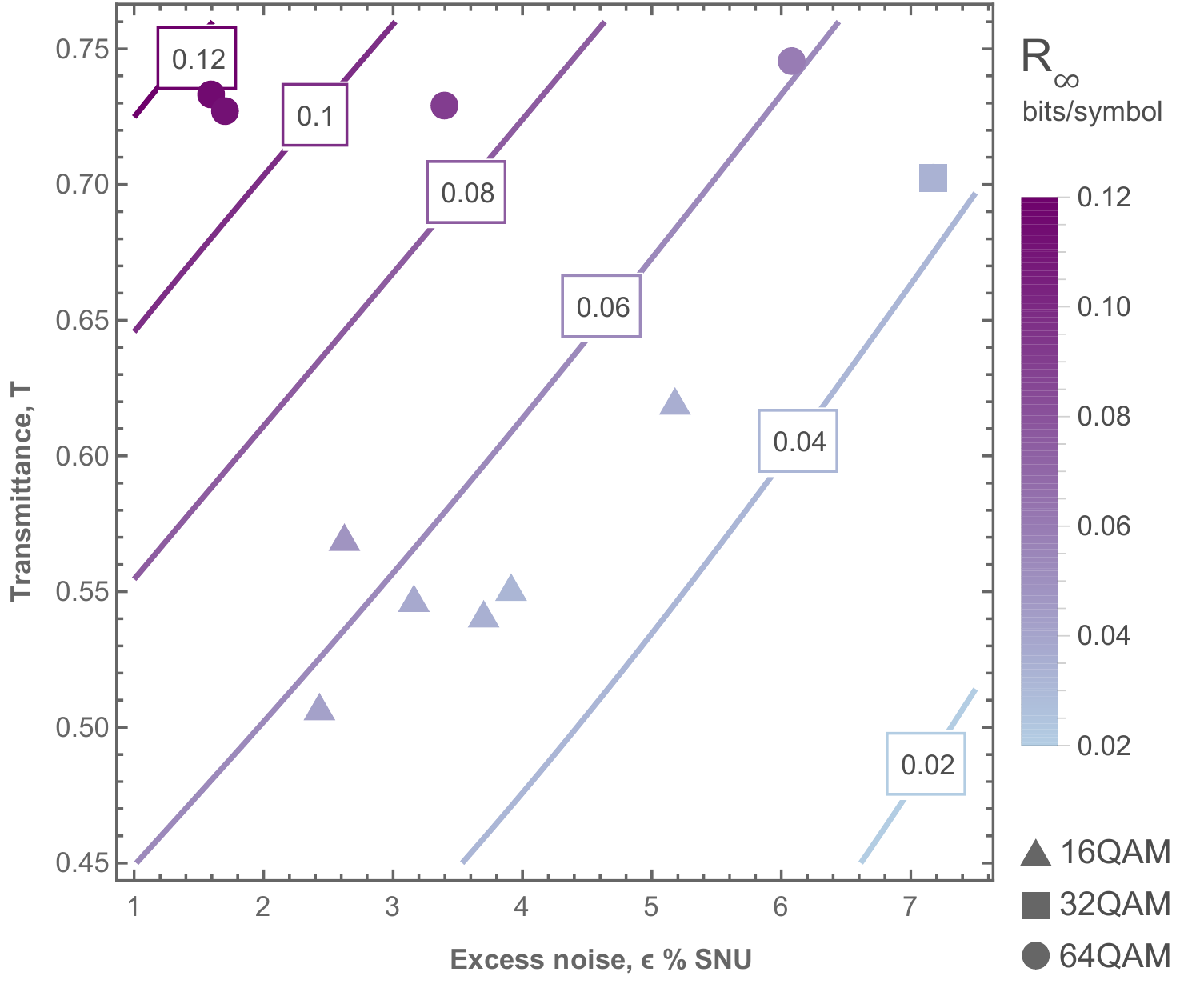}
    \caption{Dependency of the key rate on the channel loss and excess noise. Lines indicate key rate levels of GG02, and points are for experimental results.}
    \label{fig:contour}
\end{figure}

Figure~\ref{fig:contour} illustrates the impact of fluctuations in channel parameters on system performance. Slight variations in channel excess noise and untrusted loss result in noticeable differences in the asymptotic key rate (represented as points). These variations can arise due to a multitude of reasons, such as those caused by polarization or temperature fluctuations, or mechanical drift present in the optical probing of the photonic IC. Even with a small cardinality, the performance of GG02 (depicted by contour lines) can be regained when operating under conditions of low noise. This suggests that while reducing the channel excess noise and approaching the performance of GG02 protocol through low modulation variance, the system becomes more sensitive to unavoidable fluctuations in the channel parameters in practical applications.


In the absence of rigorous finite-size key evaluation techniques, we consider the significant impact of parameter estimation on the key length in this regime. Channel parameters are chosen with worst-case assumptions, ensuring that the information accessible to Eve is not underestimated. We calculate the pessimistic key rate $R_\text{finite}(T^{low},\epsilon^{up})$ (see Tab.\ref{tab:results}) within well-established confidence intervals of the Gaussian channel parameters corresponding to parameter estimation failure probability $\varepsilon^{PE}_{fail}=10^{-10}$ with 6.5 standard deviations \cite{ruppert2014long}. Additionally, since error correction precedes the parameter estimation stage, the whole raw key sequence can concurrently be used for parameter estimation and key extraction \cite{leverrier2015composable}. 

\section{Discussion}
DM CV-QKD is a technique that employs a discrete constellation of coherent states with finite cardinality. It is compatible with high-speed telecom components, making it a promising candidate for achieving ultra-high secret key rates. In this work, we have reported the fastest DM CV-QKD system that operates at a symbol rate of \SI{10}{\giga\baud}.

This record-breaking demonstration was made possible by improving the total bandwidth of the system. In particular, we designed an integrated photonic-electronic phase-diverse receiver that maintains a wide shot-noise limited bandwidth. Additionally, the transmitter bandwidth was improved through well-designed DSP, which includes a pre-emphasis filter, for quantum state preparation. These enhancements have enabled our system to generate secret keys at exceeding {\SI{0.7}{Gb\per\second}} in both the asymptotic regime and after incorporating dominant effects of the finite-size regime.

Compared to recent progress in high-rate DM CV-QKD~\cite{ wangSubGbpsKeyRate2022,pan2022experimental,milovanvcev2020spectrally,eriksson2020wavelength,tian2023high}, our demonstration not only doubles the symbol rate but also offers a compact and semi-autonomous QKD system. However, there is room for improvement in the current system, including rigorous finite-size effects estimation for non-Gaussian channel assumption, and fast implementation of information reconciliation. In addition, for enhanced system stability, the next version of our receiver will necessitate proper chip packaging. To this end, our results pave the way for ultra-high-rate QKD systems, enabling secret key-demanding applications such as real-time one-time-pad secured video encryption.

\subsection*{Data availability}
Data underlying the results presented in this paper are available from the authors upon reasonable request.
\subsection*{Acknowledgments}
 AAEH, NJ, ULA and TG acknowledge support from Innovation Fund Denmark (CryptQ, grant agreement no. 0175-00018A) and from the Danish National Research Foundation, Center for Macroscopic Quantum States (bigQ, DNRF142). This project was funded within the QuantERA II Programme (project CVSTAR) that has received funding from the European Union’s Horizon 2020 research and innovation programme under Grant Agreement No 101017733. ID acknowledges support from the project 22-28254O of the Czech Science Foundation. CB, AB, SB and XY acknowledge support from the Digital Europe project BE-QCI (No. 101091625), Research Foundation Flanders through the Research Foundation–Flanders (FWO) Weave project Squeezed Quantum prOcessing with Photonics and Electronics (SQOPE) (G092922N). We acknowledge support from the Horizon Europe framework programme Quantum Secure Networks Partnership (QSNP, No. 101080116).
\subsection*{Disclosures}
The authors declare no conflicts of interest.

\subsection*{Author contributions statement}
A.A.E.H. and C.B. contributed equally as first authors. A.A.E.H.\ designed the experiment, implemented the DSP routine, and performed the overall data processing and analysis under supervision of TG. C.B.\ designed the integrated Photonic-Electronic Chip under the supervision of X.Y.. A.A.E.H., C.B.\ and N.J.\ implemented the experimental setup, and A.A.E.H.\ and C.B.\ performed the experimental measurements. I.D.\ performed the security analysis. A.B.\ and S.B.\ performed chip characterization. A.A.E.H.\ and T.G.\ wrote the manuscript with input from I.D.\ and C.B.. A.A.E.H.\ and T.G.\ conceived the experiment. U.L.A., X.Y.\ and T.G.\ supervised the project. All authors were involved in discussions and interpretations of the results.
\bibliography{referencesCB.bib}


\bibliographyfullrefs{referencesCB.bib}

\end{document}